# 2π scaling law during inviscid fluid pinch-off and satellite droplets formation

*Dege Li, Yi Cao, Bingfang Huang, Yonghong Liu, and Yanzhen Zhang\**

[a]College of Mechanical and Electronic Engineering, China University of Petroleum (East China), Qingdao 266580, China

Pinch-off and satellite droplets formation during breakup of near-inviscid liquid bridge sandwiched between two given equal and coaxial circular plates have been investigated. The breakup always results in the formation of a spindle shape which is the precursor of the satellite droplet at the moment of pinch-off. Interestingly, the slenderness of this spindle is always bigger than 2π and always results in the formation of only one satellite droplet regardless of the surface tension and the slenderness of the liquid bridge. We predict the cone angle of this spindle formed during the pinch-off of inviscid fluids should be 18.086122158…°. After pinch-off, the satellite droplets will drift out of the pinch-off regions in the case of symmetrical short bridge, and merge again with the sessile drop in the case of unsymmetrical long bridge. We demonstrate that the velocity of the satellite droplet is consistent with a scaling model based on a balance between capillary forces and the inertia at the pinch-off region.

The breakup of liquid filaments and formation of drops, sometimes accompanied by the generation of satellite droplets, is an extremely generic nature phenomenon, such as leaky faucets, raindrop fragmentation and mist hovers over the falls. It is also related to extensively industry applications, such as inkjet printing, spray and atomization. For inkjet printing, the formation of satellite droplets is unwelcome since they tend to blur the printed patterns[1]; however, in some special applications, such as satellite droplets printing (SDP) [2, 3], they are preferable since their size is much smaller than the primary drops which have similar size with the orifice and much higher printing resolution can be realized by satellite droplets without shrinking the size of orifice. It had been demonstrated that the satellite droplets with radius only 1 or 2 micron can be actively generated by controlling the breakup of a liquid bridge and accurately conveyed by an air flow to realize printing with resolution higher than 10000 DPI, otherwise impossible for the traditional IJP techniques [2]. A comprehensive understanding of the mechanisms of liquid bridge pinch-off and satellite droplets formation is very necessary for promoting the industry applications where efforts are made either to avoid, or to utilize the satellite droplets. However, most of the attentions were put previously on the static configurations[4], recently on the dynamic breakup process [5-10] of the liquid bridge, or on the volume of the sessile drops left on the surface after pinch-off [11], rare attentions were put on the satellite droplets.

Here, the formation characteristics of the satellite droplets due to the breakup of near-inviscid liquid bridge between two given equal and coaxial circular plates were investigated. Two identical stainless steel rods with radius $R_0$ = 0.5 mm are used for horizontally holding the liquid bridge. Deionized water, sometimes with surfactant addition or mixing with ethanol for adjusting the surface tension, were used to form the liquid bridge. The breakup process was observed by an optical microscope equipped with a high speed camera with a maximum frame rate of 200 kf/s and resolution of 2 μm/pixel. The liquid was first loaded into the gap by a pipette and stretched to a length very close to its critical breakup length. After dozens seconds of waiting, breakup was finally induced by evaporation and this ensures the breakup is starting from a "static" conditions. Liquid bridges with slenderness, defined as the ratio of the bridge length $L$ to the radius of the rods $L/R_0$, ranging from 1.2 to 7.4 were studied.

The bridge was bilateral symmetrical when its slenderness is smaller than about 3.6 and asymmetrical occurs for bigger slenderness. The transition from symmetrical to asymmetrical is caused by the increasing of slenderness, but discussion of this transition is beyond the scope of this paper.

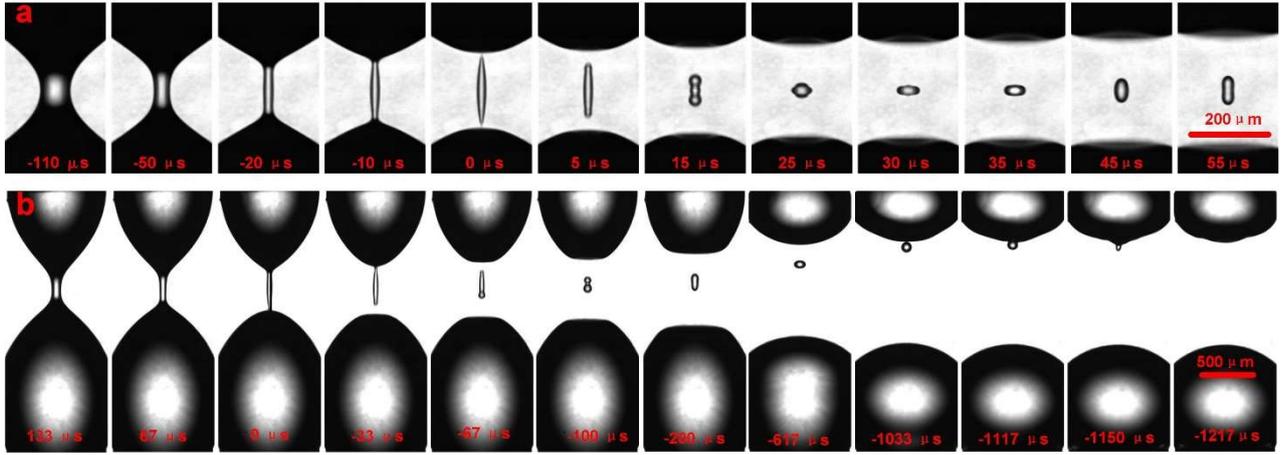

FIG. 1. Pinching off process of the (a) symmetrical and (b) asymmetric liquid bridge with slenderness 2.4 and 5.6 respectively. The surface tension $\sigma$ of the liquid used is 72 mN/m.

We found that the breakup inherits the symmetry of the liquid bridge. Namely, symmetrical bridge always leads to symmetrical breakup and asymmetrical bridge always leads to asymmetrical breakup, as shown in FIG. 1 . In the case of symmetrical bridge (FIG. 1a), the breakup is bilateral symmetrical with the same volume of liquid distributed to the two rods and only one satellite droplet pinched at the middle of the bridge.

The minimum radius of the necks $h_{min}$ scales as $\tau^{2/3}$ at both of the pinching points, where $\tau$ is the time remaining until pinch-off, regardless of the surface tension, as had been demonstrated by the pinch-off of mercury[9] which has a surface tension an order of magnitude higher than that of water. However, the experiments results showed that fluids with higher surface tension indeed breaks much faster than the fluids with higher surface tension. This can be explained by the Young-Laplace equation that the capillary pressure at the pinching off regions are proportional to surface tension. Due to the synchronous thinning of the two necks, a spindle shape which is the precursor of the satellite droplet will form at the moment of pinch-off. On the side of the sessile drops, overturning of the surface occurs. The cone angle of this spindle is measured to be about 18.1°, as expected for inviscid fluid [8]. This confirmed that for the inviscid fluid, the breakup physics of liquid bridge sandwich between two solids is essentially same with the breakup physics of jets with free surface (in the case of inkjet printing or dripping). The pinch-off is a self-similar process and is independent of the initial conditions, as confirmed by previous reported [8]. However, different with the breakup of jet with free surface, there are some notable features for the breakup of symmetry liquid bridge sandwiched between to equal size disc.

The first notable feature is the symmetry during the whole breakup process. The two pinch-off occur at the same time, and the two pinch-off point were symmetrically located at the axis of the bridge, forming a symmetrical spindle at the middle. After pinch-off, the spindle will rebound more than 10 times to dissipate the surface energy before adopting sphere shape. Experiments results showed that the period of the rebound is roughly $(\rho R_s^3/\sigma)^{1/2}$, where $\rho$ and $\sigma$ is the density and surface tension of the liquid, $R_s$ is the radius of the satellite droplet. Satellite droplets with higher surface tension rebound much faster and that with lower surface tension (FIG. 2a). Due to the symmetry, there is no net momentum left on the satellite droplets. Ideally it should keep stagnant after pinch-off due to the symmetry; however, the satellite droplets always drift out of the pinch-off region due to gravity and the air disturbance which might be caused by the breakup process itself or air flow of the environment. For the symmetrical breakup, the direction and magnitude of the velocity of the satellite droplet are random, but usually no more than 0.05 m/s.

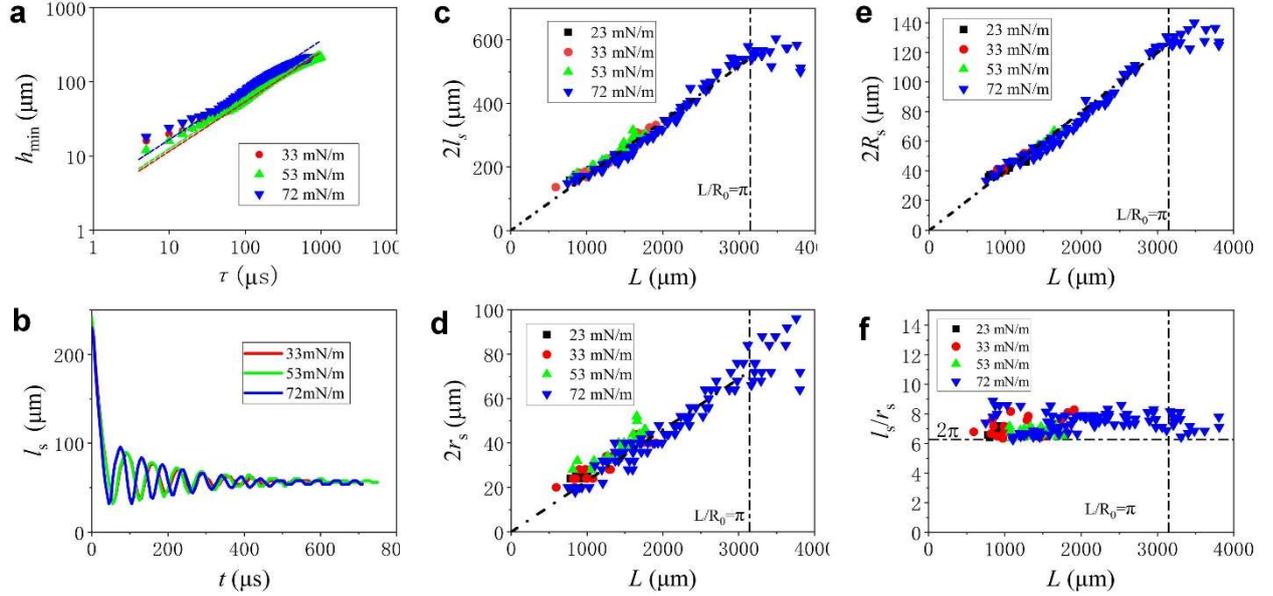

FIG. 2. Geometry characteristics during the pinch-off process of the liquid bridge. (a), neck diameter $h_{min}$ vs the time remaining until pinch-off $\tau$, (b) vibration of the satellite droplets; (c) length and (d) maximum diameter of the spindle at the moment of pinch-off versus bridge length; (e) satellite size versus bridge length; (f) slenderness of the spindle versus bridge length.

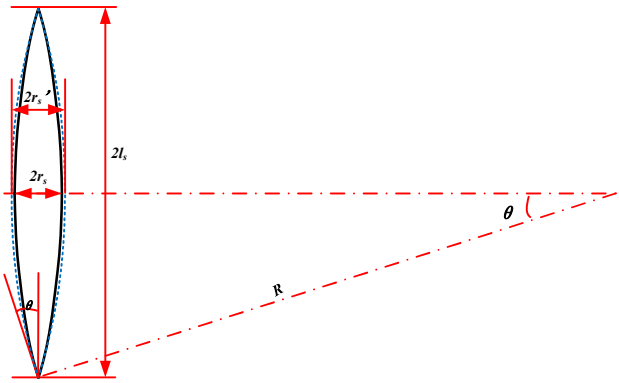

FIG. 3. Geometry illustration of the spindle shape at the moment of pinch-off. The solid line is extracted from the high speed image.

We find that the size of the spindle (both length and radius), as well as the satellite droplets size, increase linearly with the length of the liquid bridge until π (FIG. 2c-e), this might be explained by the geometric scaling effect. A notable feature of the spindle is that its slenderness of $l_s/r_s$ is always bigger than $2\pi$ regardless of the surface tension and slenderness of the bridge length, where $l_s$ is the half length of the spindle, $r_s$ is the maximum radius of the spindle (FIG. 2f). The influence of surface tension and bridge length on the slenderness of spindle is not detected by our experiments. Pursuing an analytical solution of the meridian curve of the spindle is almost impossible. Alternatively, numerical solution is feasible as has been demonstrated by Day et al[8]. Given an imaginary spindle whose meridian curve is circular arc and slenderness $l_s/r_s' = 2\pi$, as illustrated by the dotted line in FIG. 3, according to the trigonometric relation, we obtain $R^2 - l_s^2 = (R - r_s')^2$ and $\sin\theta = l_s/R$, where $r_s'$, $R$ and $\theta$ is the maximum radius, the curvature radius of the meridian curve and the cone angle of the imaginary spindle, respectively. Using the above geometry relation, the ratio between $l_s$ and $R$ can be easily obtained as $l_s/R = 4\pi/(4\pi^2 + 1)$. Therefore, the cone angle of this imaginary spindle can be expressed as $\arcsin(4\pi/(4\pi^2 + 1))$ which is 18.086122158…°, very close to 18.1° reported by Day et al [8] who numerically calculated the evolution of shape of the pinch-off region using the boundary-element method. Pursuing more decimal places of this angle is very difficult by the boundary-element method. This is because the curvature and capillary pressure at the pinching off regions are inversely proportional to radius and the Navier-Stokes governing equations become singular at the moment of breakup. For the other hand, the intrinsic error of the boundary-element method and the calculation capacity of the computer also limited the calculation accuracy.

We tend to believe this coincidence is not accidental. Over one hundred years ago, Rayleigh and Plateau established the classical theories for explaining the stability of near-inviscid liquids jet, known as the Rayleigh-Plateau Instability. According to their theory, the jet is unstable to arbitrarily infinitesimal disturbance when its slenderness is equal to or bigger than $2\pi$. Here, we found another "$2\pi$ scaling law" during the dynamic pinch-off process of near-inviscid fluid: the meridian curve of the spindle near the pinch-off point can be approximated by a spindle whose meridian curve is circular arc and slenderness is exactly $2\pi$. It seems that the "$2\pi$ scaling law" not only governing stability of the static liquid jet, but also its dynamic pinch-off process. It's well known that the notorious Navier-Stokes equation does not have analytical solution except for very special boundary conditions. Solving of the Navier-Stokes equation which governing the stability of the static inviscid jet with free surface, carried out by Plateau, is one of those exceptions. Here the "$2\pi$ scaling law" discovered from the simple geometry relationship shown in FIG. 3, indicating the pinch-off of the inviscid jet is another exception.

Due the symmetry, the curvature of the meridian curve of the spindle should be zero at the middle point and increase monotonically from the middle point to the two pinch-off points. This explains why the maximum radius of the spindle $r_s$ is smaller than the radius of the imaginary spindle $r_s'$, and why the slenderness of the spindle is always bigger than $2\pi$, as shown in FIG. 2f.

For longer liquid bridge, the meridian curve of the bridge is no longer symmetrical. The pinching off process has some notable differences with the short symmetrical bridge. In order to convenient the following discussion, we define the side where more liquid mass is distributed to as side A and the other as side B.

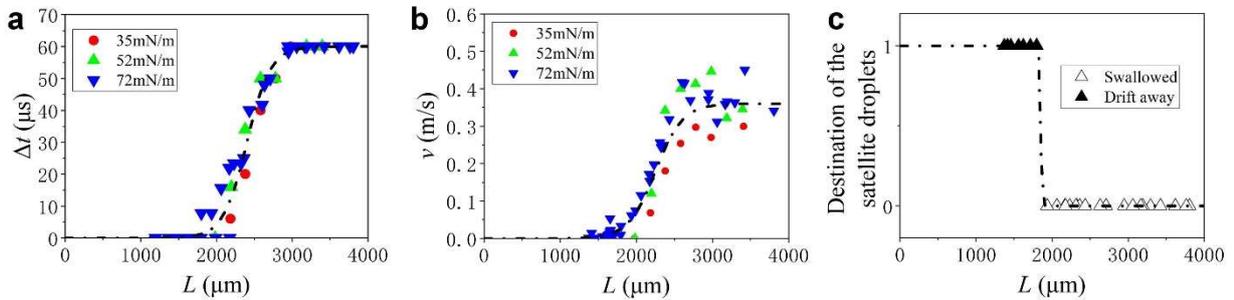

**FIG. 4.** Influence of bridge length on **a)** velocity, **b)** time difference during pinch-off and **c)** destination of the satellite droplets.

For the asymmetrical bridge, the pinch-off first occurs at the necks close to side A. There exists a time interval, $\Delta t$, between the two pinch-off. Experiments results showed $\Delta t$ increase with the increasing bridge length (FIG. 4a). The location of the two pinch-off points is closer to side B than to side A (FIG. 1b). After pinching off, the satellite droplets will get an initial velocity with directions towards side B; the magnitude of this velocity increase with increasing bridge length (FIG. 4b). This initial velocity will lead to collision with the sessile drops and the subsequent coalescence. Surprisingly, although drops bouncing at liquid surface had been extensively reported and confirmed as a generic phenomenon, satellite bouncing at surface of the sessile drops was not observed by hundreds trials of our experiments. This might be due to the small size and high speed of the satellite droplets which is unfavorable for forming the gas cushion between the satellite droplets and the surface.

As shown in FIG. 4c, depending on the velocity, with in turn is related to the slenderness of the liquid bridge, the satellite droplets have two destinations, drift out of the pinching regions or swallowed by the sessile drops. The transition of the destinations of the satellite droplets is consistent with the transition of the symmetry of the liquid bridge. Short symmetrical bridge always generate satellite that can drift out whereas long asymmetrical bridge lead to collision and swallow.

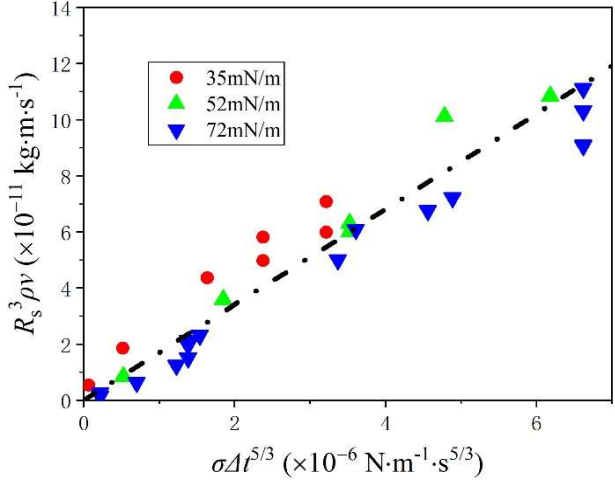

FIG. 5. Impulse-momentum scaling prediction using the same data and color code as in Fig. 2(e) and Fig 4 (a, b).

There is a time interval $\Delta t$ between the two pinch-off, therefore, the size of the two necks should be different at the same time. The kenetic of the satellite droplet should origiate from the unequal force applied on the spindle at the two necks. The total impulse $Ft$ should be equal to the momentum of the satellite droplet, where $F$ is the resultant force applied on the spindle, $t$ is the applied time of this resultant force. Quantitatively determine of applied times at the necks is difficults, nonetheless, time aplied at the two necks should have a time difference of $\Delta t$ and it is reasonable to assume $t$ as $(n+1)\Delta t$ and $n\Delta t$ at the two necks, respectively, where n is a positive integer. Therefore, the total impulse can be expressed as $Ft = \int_0^{(n+1)\Delta t} 2\pi h_{min}\sigma dt - \int_0^{n\Delta t} 2\pi h_{min}\sigma dt\ Ft$ .

Substitute the well established scaling law $h_{min} \propto \tau^{2/3}$ into this impulse equation, we obtained $Ft \propto \sigma\Delta t^{5/3}$.

Since the final momentum of the satellite droplet $mv = 4/3\pi R_s^3 \rho v$, we obtain the scaling estimate

$$\sigma\Delta t^{5/3} \propto R_s^3\rho v. \qquad (1)$$

All of the experimentally measured satellite droplet velocity, sizes and pinching off time difference from Fig. xxx are replotted in Fig. 4 versus Eq. (1). We see that the impulse-momentum scaling prediction provides an excellent estimate of the size and velocity of the satellite droplets over a wide range.

In summary, the breakup of liquid bridge with minimum stability volume sandwiched between two equal and coaxial circular plates were systematically investigated. Slenderness of the liquid bridge is the critical factor that determine the size, velocity and final destination of the satellites droplets. A spindle with slenderness always bigger than $2\pi$ will be formed at the moment of pinch-off. Geometrical analysis of the spindle indicates the dynamic pinch-off process of inviscid liquid jet is governed by a "$2\pi$ scaling law". Based on this "$2\pi$ scaling law", we made the following two predictions. First, the cone angle of the spindle should be exactly 18.086122158…° at the moment of pinch-off for inviscid fluid. Second, the Navier-Stokes equation governing the self-similar pinch-off of the inviscid fluid should have an analytical solution, similar to Rayleigh-Plateau Instability which governed the stability of liquid jet. More mathematical derivations and experimental works are still needed for revealing the physics behind the dynamic pinch-off process.

Dege Li and Yi Cao contributed equally to this work. We thank the National Natural Science Foundation of China (Grant No.: 51774316, 52075548), Taishan Scholar Program of Shandong Province (Grant No.: tsqn201909068) and Fundamental Research Funds for the Central Universities (Grant No.: 20CX06074A) for support.